\documentclass[twocolumn,aps,prb,10pt]{revtex4}
\usepackage{amsmath, bm}
\usepackage{graphicx} 
\usepackage{amssymb}
\usepackage{citesupernumber}
\usepackage{float}

\newcommand{\lp}{\left(} \newcommand{\rp}{\right)}

\newcommand{\scr}{{\mbox{\scriptsize cr}}}
\newcommand{\smicro}{{\mbox{\scriptsize micro}}}
\newcommand{\svibr}{{\mbox{\scriptsize vibr}}}
\newcommand{\sex}{{\mbox{\scriptsize ex}}}
\newcommand{\sco}{{\mbox{\scriptsize c}}}
\newcommand{\ssh}{{\mbox{\scriptsize s}}}

\newcommand{\prtl}{\partial}
\newcommand{\la}{\left\langle}
\newcommand{\ra}{\right\rangle}
\newcommand{\br}{\bm r}
\newcommand{\bn}{\bm n}
\newcommand{\bu}{\bm u}
\newcommand{\btu}{{\tilde{\bu}}}
\newcommand{\tu}{{\tilde{u}}}
\newcommand{\bd}{\bm d}
\newcommand{\bg}{\bm g}
\newcommand{\be}{\bm e}
\newcommand{\bk}{{\bm k}}

\newcommand{\hk}{\hat{k}}

\newcommand{\vg}{\vec{g}}
\newcommand{\vphi}{\vec{\phi}}

\newcommand{\tX}{{\mbox{\scriptsize X}}}
\newcommand{\tG}{{\mbox{\scriptsize G}}}

\newcommand{\scompr}{{\mbox{\scriptsize compr}}}
\newcommand{\sshear}{{\mbox{\scriptsize shear}}}
\newcommand{\sCurie}{{\mbox{\tiny Curie}}}
\newcommand{\sMF}{{\mbox{\tiny MF}}}
\newcommand{\sspin}{{\mbox{\scriptsize spin}}}


\begin{document}

\newlength{\figurewidth}
\setlength{\figurewidth}{\columnwidth}
\title{\bf \Large Stress distribution and the fragility of supercooled
  melts}
\author{Dmytro Bevzenko and Vassiliy Lubchenko \\ {\em Department of
    Chemistry, University of Houston, Houston, TX 77204-5003}}
\date{\today}
\begin{abstract}

  We formulate a minimal ansatz for local stress distribution in a
  solid that includes the possibility of strongly anharmonic
  short-length motions. We discover a broken-symmetry metastable phase
  that exhibits an aperiodic, frozen-in stress distribution. This
  aperiodic metastable phase is characterized by many distinct, nearly
  degenerate configurations.  The activated transitions between the
  configurations are mapped onto the dynamics of a long range
  classical Heisenberg model with 6-component spins and anisotropic
  couplings. We argue the metastable phase corresponds to a deeply
  supercooled non-polymeric, non-metallic liquid, and further
  establish an order parameter for the glass-to-crystal
  transition. The spin model itself exhibits a continuous range of
  behaviors between two limits corresponding to frozen-in shear and
  uniform compression/dilation respectively. The two regimes are
  separated by a continuous transition controlled by the anisotropy in
  the spin-spin interaction, which is directly related to the Poisson
  ratio $\sigma$ of the material. The latter ratio and the
  ultra-violet cutoff of the theory determine the liquid
  configurational entropy. Our results suggest that liquid's fragility
  depends on the Poisson ratio in a non-monotonic way. The present
  ansatz provides a microscopic framework for computing the
  configurational entropy and relaxational spectrum of specific
  substances.
\end{abstract}
\maketitle



If cooled sufficiently rapidly, a liquid may fail to crystallize, but
will instead remain in a metastable, {\em supercooled} state. Upon
further cooling, the relaxation times in a supercooled liquid grow
very rapidly as the mass transport becomes activated, in contrast with
the mainly collisional transport near the fusion temperature.  Because
the local structures are much longer-lived than the vibrational
equilibration times, the activated-transport regime represents a state
with a broken translational symmetry, even though the corresponding,
aperiodic structure shows no obvious distinction from a snapshot of an
ordinary, uniform liquid. (It is said the heterogeneity is
``dynamical.'')  Since the symmetry is broken {\em gradually} with
lowering the temperature - beginning with the highest frequency
motions - a transition into this ``aperiodic-crystal'' state is not
sharp, but, instead, is a soft cross-over centered at a temperature
$T_\scr$, \cite{dens_F1, LW_soft} corresponding universally to
viscosity 10 Ps or so.\cite{LW_soft, Roland} The cross-over into the
activated regime is a finite-dimensional analog of a mean-field
kinetic catastrophe of the mode-coupling theory (MCT), whereby the
motional barriers would diverge at a temperature $T_A$, even though
the configurational entropy is still extensive.\cite{MCT, MCT1}

In additional contrast to the mean-field transition at $T_A$, the
cross-over at $T_\scr$ exhibits two emerging length scales: One length
scale is the molecular length $a$ that signifies the volumetric size
of a chemically rigid unit - often called the ``bead'' - that is not
significantly perturbed during activated transport. Conversely, the
beads interact with each other weakly, comparably to the Lennard-Jones
interaction.\cite{LW_soft} The bead may be thought of as a
coarse-graining length, beyond which activated motions are largely
independent of chemical detail, but fully characterized by a single,
{\em bulk} quantity. This bulk quantity is the excess liquid entropy
relative to the corresponding crystal, usually called the
configurational entropy. The magnitude of the configurational entropy
per bead, $s_c$, directly gives the number of alternative aperiodic
configurations available to a region of a supercooled liquid
containing $N$ beads, i.e.  $e^{s_c N/k_B}$.  The bead usually
contains two-three atoms, but could be bigger for molecular liquids
containing large rigid units such as benzene. The other length scale
emerging during the crossover is the so called Lindemann length
$d_L$,\cite{dens_F1, Lindemann, L_Lindemann} which is the molecular
displacement at the mechanical stability edge. This length is nearly
universal: $d_L \simeq a/10$, and characterizes bead displacements
during transitions between distinct aperiodic packings in the
metastable, aperiodic crystal phase. One may view the crossover into
the activated liquid regime as a ``localization'' transition, whereby
the emerging metastability of local structures is signaled by a {\em
  discontinuous} transition from a uniform liquid to a state with a
non-zero force constant of the Einstein-oscillator.\cite{dens_F1} The
random first order transition (RFOT) theory utilizes this view to
analyze the activated transport in chemically distinct (non-polymeric)
fluids in a {\em unified} fashion.\cite{KTW, XW, LW_aging} (See
Ref.\cite{LW_ARPC} for a review.)  The RFOT theory predicts that in a
fully developed activated regime, the structural relaxation time is
determined solely by the configurational entropy per bead:\cite{XW,
  LW_soft}
\begin{equation} \label{tau_sc} \tau = \tau_\svibr \exp(32 k_B/s_c),
\end{equation}
where $\tau_\svibr \simeq 1$ psec is the microscopic time scale
characterizing vibrational relaxation. By Eq.(\ref{tau_sc}),
system-specific deviations from the pure Arrhenius temperature
dependence of $\tau$ result from variations in the value of the heat
capacity jump at the glass transition temperature $T_g$ per bead:
$\Delta c_p \equiv T (\prtl s_c/\prtl T)|_{T=T_g}$.  The so called
fragility index $m \equiv d(\log \tau)/Td(1/T)|_{T=T_g}$ gives a
quantitative measure of that deviation. Small and large deviations
from the Arrhenius $T$-dependence (corresponding to small and large
$m$ respectively) are often called strong and fragile
behaviors.\cite{MartinezAngell, Angell_fragility} The RFOT theory
predicts $m \simeq 34.7 \Delta c_p$,\cite{XW, LW_soft, StevensonW} in
excellent agreement with experiment.

Hall and Wolynes have put forth a simple specific model that relates
the degree of molecular connectivity to the heat capacity jump $\Delta
c_p$.\cite{HallWolynes} Establishing such connections between
local-chemistry and thermodynamics for actual substances is difficult,
however, hampering our efforts to make first principles estimates of
the configurational entropy and, ultimately, the glass-forming ability
of those substances. Indeed, since the viscosity is directly related
to the average relaxation time,\cite{Lionic}
\begin{equation} \label{eta1} \eta = \frac{2 k_B T}{\pi a d_L^2} \la
  \tau \ra \simeq 60 \: \frac{k_B T}{a^3} \la \tau \ra,
\end{equation}  
the crystal nucleus growth, which is subject to viscous
drag,\cite{Turnbull} is determined by the configurational entropy
$s_c$ per bead. The actual chemical identification of beads is
relatively straightforward in molecular, but less so in covalently
bonded compounds, where, for instance, the apparent bead often
corresponds to a non-integer fraction of a stoichiometric
unit.\cite{LW_soft, ZL_JCP} We note that these challenges are, of
course, not unique to theories of the glass transition but equally
pertain to quantitative descriptions of the ordinary liquid-to-crystal
transition. Yet perhaps in reflection of these open questions, which
we view as quantitative, many believe fragile and strong behaviors
actually have distinct mechanisms.

In an effort to develop a microscopic description of the
configurational dynamics in specific substances, here we propose and
work out several consequences of a novel quasi-continuum ansatz for
local stress distribution that implements direct interactions in the
activated regime semi-empirically, via the local elastic properties of
the material. The ansatz incorporates the possibility of local,
short-length motions that are similar in spirit to Einstein
oscillators, but are strongly anharmonic and account explicitly for
the tensorial nature of the relation between local deformation and
stress in solids.  In particular, explicitly treated is
(high-frequency) {\em shear} resistance, which is characteristic of
deeply supercooled liquids and, generally, activated liquid
transport. This fully tensorial treatment of stress may be viewed,
among other things, as a systematic improvement on self-consistent
scalar-phonon theories of aperiodic crystals.\cite{dens_F1,
  ISI:A1987G269600055}

The many-body effects that lead to the emergence of shear resistance
in liquids at high densities do not lend themselves easily to
perturbative treatments that use the uniform liquid as the reference
state. In contrast, we use a fully mechanically stable state as the
reference state, and then uniformly allow for local, short-length
anharmonic motions.  Within this ansatz, a {\em metastable}, aperiodic
frozen-in stress pattern emerges self-consistently, whereby the bead
size is identified as the typical length scale of the stress
heterogeneity.  This result implies that even in a covalently bonded
material, the concept of a bead is entirely unambiguous. The frozen-in
stress pattern is multiply {\em degenerate}, whereby the transitions
between alternative configurations can be mapped onto the dynamics of
a classical Heisenberg model with six-component spins. The six
components correspond to the independent entries in the local
frozen-in deformation tensor. The average length of the spin emerges
as an order parameter for the aperiodic-to-periodic crystal
transition. The transitions between different spin configurations are
rare events implying the corresponding liquid states are long-lived
and thus supporting the view of liquids in the activated-transport
regime as aperiodic crystals. In other words, we explicitly confirm
the non-trivial notion of the RFOT theory that aperiodic molecular
assemblies can exhibit multiple states and, at the same time, support
shear.

Already in a mean-field analysis of the spin model, we establish that
the broken-symmetry state can exhibit a range of behaviors
interpolating between frozen-in shear and uniform
compression/dilation. These two limits correspond to a 5-component
Heisenberg- and Ising-like ferromagnets respectively.  The resulting
heat capacity variations, together with the RFOT-derived intrinsic
relation between thermodynamics and kinetics, as reflected in the $m$
vs. $\Delta c_p$ relation, implies that the spin model exhibits a
broad range of fragile-to-strong behaviors during activated
rearrangements.  The heat-capacity jump per bead of the corresponding
liquid will also depend on the ultraviolet cutoff of the theory, as
specified by the bead size. We will observe that the fragility depends
on the Poisson ratio of the material, but in a complicated,
non-monotonic fashion.  In any case, in view of the apparent broad
range of strong-to-fragile behaviors exhibited in the present model,
the present analysis strongly suggests the activated dynamics have the
same mechanism in both strong and fragile liquids, consistent with the
basic picture of the RFOT theory.

A potential formal benefit of the present approach is that it is
possible to map the activated-transport liquid regime onto a spin
model on a {\em fixed} lattice, thus opening the possibility of
applying to liquids the many computational techniques developed in the
context of spin models. Importantly, this mapping may provide a first
principles basis for computing the configurational entropy for
specific substances and the rate of crystallization from the
supercooled state, since elastic constants at the length scale of a
bead, which are the input into the theory, can be determined by ab
initio methods.

\section{Microscopic ansatz for local stress distribution}
\label{Ansatz}

{\em Informal motivation.} Consider a substance that can both
crystallize and vitrify and assume for concreteness that the glass has
a greater specific volume than the corresponding crystal, as is
usually the case. A quenched melt or frozen glass of such a substance
may be thus viewed as a result of the following action: Apply negative
pressure to slowly expand the crystal until the lattice relaxes into
one of the myriad available glass structures, and then release the
negative pressure. (We assume cracking does not occur.)  The sample
will remain in the metastable, glass state; the eventual transition to
the lower free energy, crystal state will occur by nucleation and is
subject to the surface tension between the two phases. In a standard
fashion, the surface tension implies that the dependence of the free
energy on the extent of dilation exhibits a non-concave
portion.\cite{RowlinsonWidom} (By ``concave,'' we will consistently
mean concave-{\em up}.)

\begin{figure}[t]
\centering
\includegraphics[width=0.7\figurewidth]{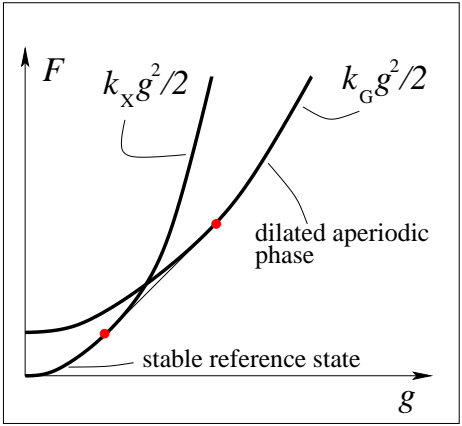}
\caption{\label{heuristic} \small Illustration of the notion of an
  aperiodic metastable state as a dilated crystal that was allowed to
  relax.} 
\end{figure}

We can see qualitatively how the non-concavity arises by estimating
the free energy of the solid using a simple {\em non-linear}
Einstein-oscillator model. Suppose the configurational partition
function of each oscillator is $\int_{g}^\infty d^3\br e^{-k r^2/2 k_B
  T}$, where we incorporate the non-linearity qualitatively by
assuming a hard lower cut-off in the configurational integral, at a
displacement $g$.  $k$ is the force constant in the harmonic limit.
Denote the force constants in the crystal and the expanded state as
$k_\tX$ and $k_\tG$ respectively. Clearly, $k_\tG < k_\tX$. We sketch
the free energies of the states as functions of $g$ in
Fig.\ref{heuristic}, taking into account that the curve corresponding
to the expanded state is shifted upward from the crystal state by the
free energy cost of expansion. According to Fig.\ref{heuristic}, under
sufficient negative pressure (essentially equal to the negative slope
of the common tangent of the two curves), the crystal will indeed
begin converting into a collection of non-linear oscillators.  The
vast majority of the configurations of these oscillators are of course
aperiodic, implying the metastable phase is aperiodic too. This
qualitative free energy graph tentatively confirms the view above of a
quenched-melt/frozen-glass as a sufficiently expanded crystal that was
allowed to relax. The graph also suggests an appropriate order
parameter for the transition, namely the amplitude of
short-wavelength, strongly anharmonic motions.

In writing a {\em formal} ansatz for a solid that allows short length
anharmonic motions, we first point out that local mechanical
instabilities are perfectly compatible even with macroscopic rigidity,
examples including orientational glasses\cite{LoidlARPC} and
ferroelectrics of order-disorder type.\cite{PhysRevLett.3.412,
  PhysRevLett.67.3412, PhysRevB.67.024114} The latter systems consist
of rigid, periodic scaffolds where a {\em subset} of local motions
could be marginally stable (resulting in a vanishing frequency of
transverse optical phonons\cite{PhysRevLett.3.412}) or even
multistable. In contrast, we wish to consider here a situation where
{\em all} short length motions are allowed to be multistable. The
resulting material may or may not be macroscopically stable, as will
be determined self-consistently.  We will limit ourselves to a
dynamical regime, in which relative distortions of individual bonds
during lattice reconfigurations are only within 10\% or
so,\cite{dens_F1, Lindemann, L_Lindemann} as appropriate both for
activated transport and stable crystals. Such small distortions imply
the harmonic approximation for local stress is semi-quantitative. Take
the Landau local free energy density as a function of the local
deformation profile $u_{ij}$: $f = \frac{1}{2} u_{ij} \,
\Lambda_{ijkl} \, u_{kl}$, where $\Lambda_{ijkl}$ is the standard
elastic tensor and the Einstein's summation convention for {\em Latin}
indeces is implied.\cite{LLelast} ($u_{ij} \equiv \frac{1}{2}(\prtl
u_i/\prtl x_j + \prtl u_j/\prtl x_i)$, $i, j = 1, 2, 3$, where $\bu$
is the local displacement relative to a steady-state reference state.)

Decompose the deformation tensor into long- and short- wave length
components: $u_{ij} = u_{ij}^< + d_{ij}$, where $u_{ij}^<$ only
includes Fourier components with wave-vectors within a certain cut-off
surface in $k$-space, while $d_{ij}$ contains only Fourier components
outside of the cut-off surface. Let us evaluate the free energy of the
system subject to a constraint on the self-energy of the local motions
$d_{ij} \Lambda_{ijkl} d_{kl} = g^2$, thus forcing the short-length
motions to be essentially anharmonic Einstein oscillators. If the
cut-off surface in the Fourier space is generically at $k \simeq
k_\smicro$, the anharmonic oscillators fill the space, approximately
one per volume $(\pi/k_\smicro)^3$, but not necessarily in a periodic
fashion. The detailed lattice in which the oscillators are arranged
depends on the precise shape of the cut-off surface.  Since the
presence of strongly anharmonic degrees of freedom generally requires
expanding the lattice from the reference state, owing to steric
repulsion, we must include in the full free energy of our solid a
penalty for increasing $g$, which we denote as $F_\sex(g)$. The
precise form of this function is not essential for the majority of our
conclusions; clearly it is a monotonically increasing, {\em concave}
function of $g$, to ensure the compressibility is positive.  As a
result, the full free energy is:
\begin{equation}\label{Ffull}
  F(g) = F_0(g) + F_\sex(g),
\end{equation}
where the free energy of the ansatz $F_0$ is computed by summing over
all configurations of local stress at temperature $T \equiv 1/k_B
\beta$:
\begin{equation} \label{Fug} e^{-\beta F_0(g)} = \sum_{\{\bu(\br),
    \bd(\br)\}} e^{-\beta \int \! \! dV f(\br)} ,
\end{equation}
subject to the constraint
\begin{equation} d_{ij}(\br) \Lambda_{ijkl} d_{kl}(\br) = g^2(\br).
\end{equation}
The local free energy density $f(\br)$ of a particular stress
configuration at site $\br$ is
\begin{equation} \label{fr}
  f(\br) = \frac{1}{2} (u_{ij} + d_{ij}) \,
  \Lambda_{ijkl} \, (u_{kl} + d_{kl}).
\end{equation}
We have dropped the superscript ``$<$'' from the long-wavelength,
``acoustic'' component of the total deformation $u^<$, to simplify
notation. For the sake of argument, we will assume the order parameter
is uniform: 
\begin{equation}
  g^2(\br) = g^2. 
\end{equation}
Formal extension to a non-uniform $g(\br)$ is straightforward.

In the following, it will be often convenient to present the six
independent entries of the symmetric tensors $u_{ij}$ and $d_{ij}$ as
6-component vectors, which is sometimes referred to as the Voigt (or
Voigt-Mandel) notation.\cite{Voigt, Mandel} Here we choose the
specific realization of the Voigt notation, in which the tensor
$\Lambda$ is a second rank (positively defined) {\em tensor} of size
$6\times6$. Upon a suitable linear transformation, see Appendix
\ref{Voigt}, we can formally rewrite Eq.(\ref{fr}) as
\begin{equation} \label{Fugvec} f(\br) \equiv
  f_{\vphi,\vg}(\br) = \frac{1}{2} ( \vphi + \vg)^2,
\end{equation}
where
\begin{equation} \label{phidef} \phi^2 \equiv \sum_{\alpha=1}^6
  \phi_\alpha^2 = u_{ij} \, \Lambda_{ijkl} \, u_{kl},
\end{equation}
and
\begin{equation} \label{gdef} g^2 \equiv \sum_{\alpha=1}^6 g_\alpha^2
  = d_{ij} \, \Lambda_{ijkl} \, d_{kl}.
\end{equation}
Throughout the article, we will consistently indicate 6-component
vectors with arrows: $\vg = \sum_{\alpha=1}^6 g_\alpha
\vec{e}_\alpha$, where $\vec{e}_\alpha \vec{e}_\beta = \delta_{\alpha
  \beta}$; and three component vectors with bold symbols: $\bu = u_i
\be_i$, where $\be_i \be_j = \delta_{ij}$. We will not apply the
Einstein summation convention to Greek indeces.

Even though we have written $\vg(\br)$ (or $d_{ij}(\br)$) as a
continuum field, it is understood that $\vg(\br)$ actually stands for
a discrete vector located at the site of the lattice of the non-linear
Einstein oscillators that is closest to the point $\br$ in
space. Switching between discrete summation over lattice sites and
continuous integration over space is straightforward: If the volume
occupied by a single vector $\vg$ is $a^3$ - implying $k_\smicro
\simeq \pi/a$ - we may interchange the summation and integration
according to $a^3 \sum \leftrightarrow \int dV$.  Since vectors $\vg$
correspond to 6 degrees of freedom, the region corresponding to one
vector $\vg$ must contain at least {\em two} atoms.  $\vphi(\br)$ (or
$u_{ij}(\br)$) is of course, too, an approximation to acoustic modes
of a {\em discrete} lattice, at wave-lengths exceeding
$2\pi/k_\smicro$. Note that the number of these acoustic modes is
three per vector $\vg(\br)$, corresponding to the three translational
degrees of freedom of the vector-containing region.

For the sake of argument we will consider the simplest case of an
isotropic reference state:
\begin{equation} \label{Liso}
\Lambda_{ijkl} = \lambda \delta_{ij}
\delta_{kl} + \mu (\delta_{ik} \delta_{jl} + \delta_{il}
\delta_{jk}), 
\end{equation}
where $\lambda$ and $\mu$ are the Lam\'{e} coefficients.\cite{LLelast}
$\mu$ is also called the shear modulus.  In the isotropic case, the
components of vectors $\vphi$ and $\vg$ have a particularly lucid
meaning.  One component - called ``dilatational'' - corresponds to
uniform compression:
\begin{equation} \label{compr}
  g_1^2 = K d_{ii}^2, 
\end{equation}
and the rest five - called ``isochoric'' - to pure shear:
\begin{equation} \label{shear} \sum_{\alpha=2}^6 g_\alpha^2 = 2 \mu
  (d_{ij} - \frac{1}{3} \delta_{ij} d_{ll})^2,
\end{equation}
where $K \equiv [\lambda + (2/3) \mu]$ is the usual bulk
modulus.\cite{LLelast} 

\section{Glass as a Frozen-In, Multiply Degenerate Stress Pattern}
\label{Glass}

We inquire about the possibility of (meta-stable) states with non-zero
$g$ by testing for non-concavity of $F(g)$ from Eq.(\ref{Ffull}). The
latter peculiarity is only possible if $F_0(g)$ has an inflection
point. An analytic evaluation of $F_0(g)$ appears difficult (except in
the mean-field limit, see next Section). Despite the simple bilinear
form in Eq.(\ref{fr}), the functional integration over $\bu(\br)$ in
Eq.(\ref{Fug}) at a fixed configuration $d_{ij}(\br)$ could not be
performed by simply changing variables $\bu = \bu' - \bd$ and
performing a Gaussian integration with respect to the new variable
$\bu'$: In contrast to the purely harmonic, long-wavelength motions
$\bu$, the new displacement field $\bu'$ would now contain
short-wavelength motions, which, by construction, are anharmonic.

It will suffice for our purposes here to extract the small and large
$g$ asymptotics of $F_0(g)$, which one may accomplish by first
averaging the Boltzmann weight $e^{-\beta \int \!  \! dV f(\br)}$ from
Eq.(\ref{Fug}) over the {\em directions} of the $\vg$ vectors on each
site. By the aforementioned prescription for interchange between
discrete summation over the oscillator sites and integration in space,
we break up the integral in the exponent into a sum $\sum_i a^3
f(\br_i)$. Note the formula $\la \exp(\vec{x} \vec{y}) \ra =
I_{m/2-1}(xy) (xy/2)^{1-m/2} \Gamma(m/2)$, where the averaging $\la
\ldots \ra$ is with respect to the mutual angle between two
$m$-component vectors $\vec{x}$ and $\vec{y}$ (see
Ref.\cite{PhysRevE.68.036115} or formula 9.6.18 of Ref.\cite{AS}).
Here, $I_\nu(x)$ is the modified Bessel function of the first kind of
order $\nu$ and $\Gamma$ is the gamma function.\cite{AS} Using this
formula, we average $e^{-\beta a^3 \vec{\phi} \vec{g}}$ over the
directions of $\vec{g}$ on each site and obtain the following free
energy at site $\br$, per non-linear oscillator:
\begin{equation} \label{Fphig} f_{\phi, g}(\br) = \frac{1}{2}
  [\phi^2(\br) + g^2] - \frac{1}{\beta a^3} \ln \frac{I_2[\beta a^3
    \phi(\br) g]}{[\beta a^3 \phi(\br) g]^2},
\end{equation}
up to an additive constant. The subscripts $\phi, g$ at $f_{\phi,
  g}(\br)$ above signify that the latter function is derived from the
original free energy density $f(\br)$ from Eq.(\ref{Fugvec}), but now
depends only on the absolute values of the 6-component vectors
$\vec{\phi}$ and $\vec{g}$ characterizing the long- and
short-wavelength distortions respectively.  We note the expression
(\ref{Fphig}) represents the exact free energy of our ansatz, up to
the small numerical ambiguities related to the detailed shape of the
cut-off surface in the Fourier space and the lattice of the spins.

To extract the small $g$ behavior of $F_0(g)$, we note that at small
values of either field, expression (\ref{Fphig}) yields
\begin{equation} \label{Fphigapp} f_{\phi, g}(\br) \xrightarrow{\phi g
    \rightarrow 0} \frac{1}{2}\phi^2(\br) \left(1 - \frac{g^2 \beta
      a^3}{6}\right) + \frac{1}{2} g^2,
\end{equation}
implying that the sole effect of the presence of the (small)
short-wavelength anharmonic displacement of magnitude $g$ is a
renormalization of all acoustic frequencies downwards by a factor of
$(1 - g^2 \beta a^3/6)^{1/2} \simeq (1 - g^2 \beta a^3/12)$. Since the
partition function of a classical harmonic oscillator with frequency
$\omega$ is $(k_B T/\hbar \omega)$, this reduction in the acoustic
frequency results in a free energy shift of $-k_B T (g^2 \beta
a^3/12)$ per phonon. As mentioned earlier, the number of the acoustic
modes is three times the number of vectors $\vg$, yielding that at
small $g$, $f(g) \simeq g^2/2 - 3 (k_B T/a^3) (g^2 \beta a^3/12) =
g^2/4$.  Thus at small $g$, the second derivative of $F_0(g)$ is
positive.

According to the asymptotic expansion of the Bessel function,
$I_\nu(x) \propto e^x/x^{1/2}$,\cite{AS} the two leading terms in the
large $g$ limit of Eq.(\ref{Fphig}) are given by
\begin{equation} \label{FphigL} f_{\phi, g}(\br) \xrightarrow{g
    \rightarrow \infty} \frac{1}{2} [|\vec{\phi}(\br)| - g]^2 +
  \frac{5}{2} \frac{k_BT}{a^3}\ln g.
\end{equation}
In contrast with Eq.(\ref{fr}), the $g$-dependent portion of the free
energy is now uniform in space and thus no longer has short-wavelength
components. As a result, we can always shift the $\bu$ variable to
eliminate the $g$ dependence from the integrand in the partition
function corresponding to the free energy (\ref{FphigL}) so as to make
the expression a pure quadratic in the new variable $\phi'$: $\bu =
\bu_0 + \bu' \Rightarrow \vec{\phi} = \vec{\phi}_0 + \vec{\phi}'$
where $|\vec{\phi}_0| =g$. (This latter reference state physically
corresponds to a uniform distribution of stress (not displacement!) at
free energy density $g^2/2$.) The resulting integration is $\phi'$
becomes asymptotically Gaussian in the large $g$ limit, despite the
non-analytical dependence on $\vec{u}$ in Eq.(\ref{FphigL}). As a
result, $f_g(g) \xrightarrow{g \rightarrow \infty} (5/2) (k_B T/a^3)
\ln g$, i.e. the second derivative of $F_0(g)$ becomes negative at
large $g$, implying $F_0(g)$ has an inflection point at an
intermediate value of $g$.

We thus conclude from the analysis of the asymptotic behavior of the
free energy $F_0(g)$ that with a suitable form of $F_\sex(g)$, such as
$F_\sex(g) \propto g^2$, it is possible for the full free energy
$F(g)$ to have a non-concave portion, also consistent with the
mean-field solution in Section \ref{MF}, see Fig.\ref{FgMF}. We
reiterate that the presence of the non-concave portion means the
system exhibits a metastable state at a {\em non-zero} value of
$g$. The latter quantity therefore can used as an order parameter for
a transition between the lowest free energy, fully stable state and
the metastable state at non-zero $g$.

\begin{figure}[t]
\centering 
  \includegraphics[width=0.8\figurewidth]{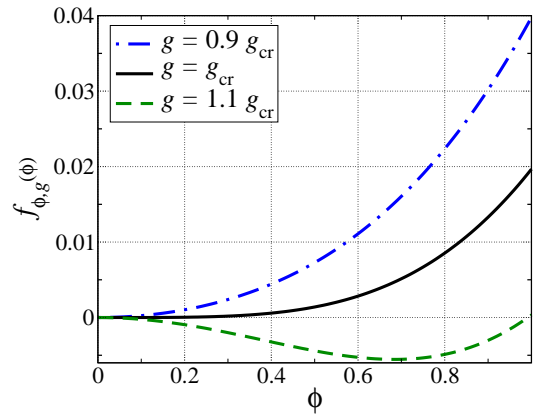}
  \caption{\label{Fphi} \small Free energy profiles for local elastic
    strain at three representative values of the anharmonic
    displacement $g$, from Eq.(\ref{Fphig}). Above the critical value
    $g_\scr$ from Eq.(\ref{gcrit}), the lattice freezes in an
    aperiodic pattern at wave-length $2\pi/k_\smicro$.}
\end{figure}

The free energy of our ansatz, Eq.(\ref{Fphig}), helps understand the
origin of the peculiar behavior of $F_0(g)$. Indeed, at large enough
magnitude of anharmonic displacement, $g$, the equilibrium value of
the acoustic displacements becomes non-zero.  According to
Eq.(\ref{Fphigapp}) this transition occurs at a critical value:
\begin{equation} \label{gcrit} g_\scr^2 = \frac{6 k_B T}{a^3}.
\end{equation}
We show the $\phi$ dependence of the free energy from Eq.(\ref{Fphig})
in Fig.\ref{Fphi}, for several representative values of $g$. Note that
the critical value of the excess free energy at a single vector, $a^3
g_\scr^2/2 = 3 k_B T$, is the total energy of a single atom in a
harmonic lattice.

In the metastable phase at $g > 0$, the thermodynamic potential from
Eq.(\ref{Fphig}) is dominated by aperiodic configurations of $\vg
\,$'s, implying the modulation of the harmonic field is aperiodic,
too.  This potential becomes particularly vivid in the $g^2/k_B T
\rightarrow \infty$ limit, i.e. $(g - |\vphi|)^2/2 \equiv [g - (u_{ij}
\, \Lambda_{ijkl} \, u_{kl} )^{1/2}]^2/2$.  This form means that the
aperiodic phase corresponds to a frozen-in stress pattern at half-wave
length $a = \pi/k_\smicro$, which we must thus associate with the bead
size of the RFOT theory.  Consistent with this identification, the
form $[g - (u_{ij} \, \Lambda_{ijkl} \, u_{kl})^{1/2} ]^2$ is similar
to broken-symmetry functionals that arise in systems with
self-generated disorder.\cite{SchmalianWolynes2000,
  PhysRevLett.75.2847} Importantly, the free energy (\ref{Fphig}) is
an equilibrium quantity, implying the metastable aperiodic state
corresponds to a melt {\em above} the glass transition.

The aperiodic phase exhibits many distinct states separated by
activation barriers. To see this structural degeneracy explicitly, we
first integrate out the acoustic deformation in Eq.(\ref{Fug}) at a
fixed $d_{ij}(\br)$ configuration. Analogous calculations have been
performed, for instance, by Grannan at el.\cite{PhysRevB.41.7784,
  Grannan_thesis}, who have considered defect-phonon couplings in the
form of $Q_{ij} u_{ij}$.  Using their result or by straightforward
path-integration in $k$-space (see Appendix \ref{K}), we obtain the
following effective Hamiltonian that couples the non-linear
oscillators, in the tensorial and Voigt notations respectively:
\begin{eqnarray} \label{H} {\cal H_\sspin} &=& - \sum_{m<n} a^3
  K_{ijkl}(\br_m - \br_n) \: d_{ij}^{(m)} d_{kl}^{(n)} \nonumber \\
  &\equiv& - \sum_{m<n} \sum_{\alpha, \beta =1}^6 a^3 J_{\alpha
    \beta}(\br_m - \br_n) \: g_\alpha^{(m)} g_\beta^{(n)},
\end{eqnarray}
where the double $(m<n)$ sums are over all bead pairs and the
spin-spin coupling is the Fourier transform:
\begin{equation} \label{Kijkl} K_{ijkl}(\br) \equiv \int_0^{k_\smicro}
  \frac{d^3(\bk a)}{(2\pi)^3} \cos(\bk \br) \tilde{K}_{ijkl},
\end{equation}
of the following tensor:
\begin{equation} \label{KijklFourier} \tilde{K}_{ijkl} = \frac{\hk_p
    \hk_q}{\mu} \left( \Lambda_{ijmp} \Lambda_{klmq} - \frac{\lambda
      +\mu}{\lambda+2\mu} \Lambda_{ijmp}\Lambda_{klnq}\hk_m\hk_n
  \right),
\end{equation}
where $\hat{\bk} \equiv \bk/k$ and the tensor $\Lambda_{ijkl}$ is from
Eq.(\ref{Liso}). The Fourier image $\tilde{J}_{\alpha \beta}(\bk)$ of
the interaction $J_{\alpha \beta}(\br)$ from Eq.(\ref{H}) is a
dimensionless $6 \times 6$ matrix, which means here that the
interaction between the spins scales as $1/r^3$ at large distance,
similarly to the usual dipole-dipole interaction.

We note that Hamiltonians similar to that in Eq.(\ref{H}) have been
employed in the context of configurational
glasses.\cite{PhysRevB.41.7784, Yu_elastic} Those treatments were
restricted to defects of the type $Q_{ij} \propto [n_i n_j -
(1/3)\delta_{ij}]$, where $\bn$ is a {\em three}-component unit vector
corresponding to the orientation of a polar group.  Grannan at
el.\cite{PhysRevB.41.7784} point out that magnets of the type in
Eq.(\ref{H}) are frustrated even on periodic lattices.

The classical Heisenberg-like ferromagnet from Eq.(\ref{H}) exhibits
many distinct states separated by barriers, below its Curie
temperature. (See next Section for the mean-field phase-diagram of the
model.) These distinct states correspond to distinct configurations in
the corresponding quenched liquid. The bead motions during transitions
between these distinct states correspond to rotations of the
six-component, classical ``spins'' $\vec{g} \,^{(m)}$. The relative
displacement of two beads, say $m$ and $n$, during a transition is
related to the transition-induced change in the vector $(\vec{g} \,
^{(m)} - \vec{g} \, ^{(n)})$. The relation, however, is not obvious
because of the tensorial nature of the frozen-in stress resulting in
vectors being six-component. As in any magnet below its ordering
transition, {\em typical} transitions between distinct metastable
states of the spin system from Eq.(\ref{H}) will involve the more
spins the lower the temperature. Such bigger cooperative regions are
necessary because the density of states for a region of fixed size
decreases dramatically for lower energies. Specifically, according to
the RFOT theory, the cooperativity size in the present case should
increase as $1/s_c^{2/3}$,\cite{KTW} where $s_c$ is the entropy per
bead (spin). Consistent with the cooperativity, the individual vectors
rotate little relatively to each other so as to minimize local strain,
implying the beads will move nearly harmonically most of the time. The
transition is, nevertheless, strongly anharmonic, as reflected in a
high free energy barrier for the reconfiguration. An appropriate
progress coordinate is the number of beads that have already
moved.\cite{LW_aging}

We conclude this Section by pointing out that the obtained mapping
between activated liquid transport and the dynamics of the 6-spin
Hamiltonian from Eq.(\ref{H}) is not exact, of course, but
nonetheless, does capture the essential features of the
activated-transport regime in liquids. Systematic improvements, such
as using a more flexible functional form for the non-linear
oscillators, including higher order multipole terms in the spin-spin
interaction etc., are possible, however will result in largely
quantitative corrections.

\section{Mean-Field Analysis of the Spin Model}
\label{MF}

It is straightforward to show that the Fourier image
$\tilde{J}_{\alpha \beta}(\bk)$ of the interaction $J_{\alpha
  \beta}(\br)$ from Eq.(\ref{H}) has the following simple property:
$\tilde{J}^2 = \tilde{J}$ (i.e. $\sum_\beta \tilde{J}_{\alpha \beta}
\tilde{J}_{\beta \gamma} = \tilde{J}_{\alpha \gamma}$), and that the
trace of the $\tilde{J}$ matrix equals three for any value of the
wave-vector $\bk$. These two notions imply the matrix $\tilde{J}$ has
two eigenvalues: $1$ and $0$, both triply degenerate. For comparison,
the analogous matrix for the regular {\em elastic} dipole-dipole
interaction between 3-component spins, $\hk_i \hk_j$ (see Appendix
\ref{K}), has eigenvalues $1$, $0$, $0$. For the {\em electric}
dipole-dipole interaction, the eigenvalues are $-1, 0, 0$. We thus
conclude that in the sense that electric dipoles form an {\em
  anti}-ferromagnet (for not too oblong samples)\cite{PhysRev.70.954}
the Hamiltonian from Eq.(\ref{H}) corresponds to a ferromagnet. (This
difference can be understood as stemming from the notion that in
phonon-mediated interactions, it is {\em like}-charges that attract,
in contrast with the Coulomb law.)

In a mean-field limit, when the interaction $J_{\alpha \beta}(\br)$
from Eq.(\ref{H}) does not depend on the mutual separation $\br$
between spins, the Hamiltonian (\ref{H}) indeed becomes a simple
ferromagnet.  This limit may be formally implemented by taking the
$k_\smicro \rightarrow 0$ limit in Eq.(\ref{Kijkl}) while
renormalizing the integrand so that the total energy of the system is
finite and scales with the total number of the 6-spins. Under these
circumstances, the $k$-integration in Eq.(\ref{Kijkl}) reduces to
angular averaging of the $\tilde{K}$ tensor. Upon the averaging, as
outlined in Appendix \ref{K}, the $\tilde{K}$ tensor becomes a
diagonal matrix with eigenvalues that turn out to be dimensionless
combinations of the microscopic elastic constants: one eigenvalue
corresponding to uniform compression: $K/(\lambda + 2 \mu)$, and a
five-fold degenerate eigenvalue corresponding to pure shear:
$(2/5)(K+2\mu)/(\lambda+2\mu)$. In other words, in the mean field
limit, the the dilatational components of local stress on one site do
not interact with the isochoric components of local stress on another
site.

As a result, the mean-field limit of the Hamiltonian (\ref{H}) is:
\begin{equation} \label{HMF} {\cal H_\sMF} = - \frac{J_\scompr}{2N}
  \left[ \sum_m^N s_1^{(m)} \right]^2 - \frac{J_\sshear}{2N}
  \sum_{\alpha=2}^6 \left[\sum_m^N s_\alpha^{(m)} \right]^2,
\end{equation}
where $\vec{s} \, ^{(m)} \equiv \vec{g}^{(m)}/g$ are unit vectors and
denominators $2N$ were introduced for convenience.  Note that the
coupling constants $J$ above are proportional to $g^2$, according to
Eq.(\ref{H}). The numerical value of the $J_\scompr/J_\sshear$ ratio
is equal to the ratio of the two distinct eigenvalues of the
angular-averaged $\tilde{K}$ tensor:
\begin{equation} \label{JJ} \frac{J_\scompr}{J_\sshear} = \frac{5}{2}
  \: \frac{K}{K + 2 \mu} \equiv \frac{5}{2} \:
  \frac{1+\sigma}{4-5\sigma} \equiv \frac{5}{2} \: \frac{3 - 4
    c_t^2/c_l^2}{3 + 2 c_t^2/c_l^2}
\end{equation}
This ratio varies between $0.88$ and $1.6$ for non-metallic glasses
surveyed by Novikov and Sokolov.\cite{PoissonFrag,
  NovikovSokolovKisliuk} Here, $\sigma$ is the Poisson ratio, $c_l$
and $c_t$ are the longitudinal and transverse speeds of sound.  Bigger
values of the $J_\scompr/J_\sshear$ ratio correspond to a greater
Poisson ratio and smaller $\mu/K$ ratio, i.e. to a lower shear modulus
relative to the bulk modulus.  We thus observe that the mean-field
Heisenberg model (\ref{HMF}) has anisotropic couplings, the anisotropy
directly related to the Poisson ratio of the material. Remarkably, the
purely isotropic case $J_\scompr/J_\sshear = 1$ (corresponding to
$\sigma = 1/5$, $c_t^2/c_l^2 = 3/8$) falls within the experimental
range, the implications to be discussed at the end of the article.

\begin{figure}[t]
\centering 
\centering
\includegraphics[width= \figurewidth]{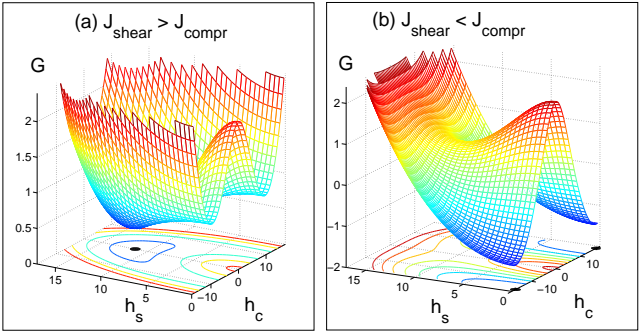}
\caption{\label{Ghh} \small Free energy from Eq.(\ref{G})
  corresponding to the elastic constants of (a) silica
  ($J_\scompr/J_\sshear \simeq 0.88$) and (b) salol
  ($J_\scompr/J_\sshear = 1.5$), corresponding to a high and low
  values of the shear modulus, relative to the bulk modulus. Cases (a)
  and (b) exhibit frozen-in shear and uniform compression/dilation
  respectively. Solid circles at bottom plane denote the locations of
  minima.}
\end{figure}

The mean-field Hamiltonian (\ref{HMF}) can be solved in a standard
fashion by a Hubbard-Stratonovich transformation, so that the
partition function is given the following six-dimensional integral, up
to a multiplicative constant:
\begin{eqnarray} \label{Z} Z_\sMF &=& \sum_{\{\vec{s}^{(m)}\}}\int
  \Pi_{\alpha=1}^6 dh_\alpha
  \exp\left\{- N \frac{ h_1^2 }{2\beta J_\scompr} \right. \nonumber \\
  &-& N \sum_{\alpha=2}^6 \frac{ h_\alpha^2}{2\beta J_\sshear} +
  \left. \sum_{\alpha=2}^6 h_\alpha \sum_m^N s_\alpha^{(m)} \right\},
\end{eqnarray}
where the sum in front of the integral denotes averaging with respect
to spin orientations. (Eq.(\ref{Z}) is a minor variation on the mean
field solution of the isotropic Heisenberg model for an arbitrary
number of vector-components, which can be found, for instance, in
Ref.\cite{PhysRevE.68.036115}) Upon the angular averaging (see the
derivation of Eq.(\ref{Fphig})), the integration can be done by
steepest descent. In the leading order, the partition function is
equal to $e^{-\beta N G_0}$, where $G_0$ is the minimum value of the
following free energy:
\begin{equation} \label{G} \frac{G}{N} = \frac{h_\sco^2}{2\beta
    J_\scompr}+ \frac{h_\ssh^2}{2\beta J_\sshear} - \ln \frac{I_2
    (\sqrt{h_\sco^2+h_\ssh^2})} {h_\sco^2+h_\ssh^2},
\end{equation}
with respect to the variables $h_\sco$ and $h_\ssh$ (c.f. Eq.(2.3) of
Ref.\cite{PhysRevE.68.036115}).  $h_\sco \equiv h_1$ gives the
effective field on each spin in the direction of uniform
compression/dilation; it can be of either sign.  $h_\ssh \equiv
(\sum_{\alpha=2}^6 h_\alpha^2)^{1/2}$ is the {\em magnitude} of the
total field in the direction of pure shear and can be only
non-negative, of course. The free energy from Eq.(\ref{G}) is graphed
in Fig.\ref{Ghh} for two distinct values of anisotropy, at a
temperature below the Curie temperature.

\begin{figure}[t]
\centering 
\includegraphics[width= 0.7 \figurewidth]{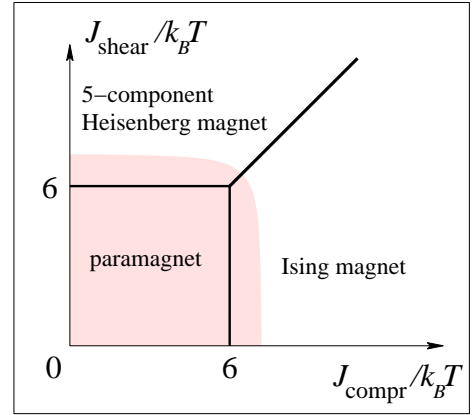}
\caption{\label{PhaseDiagram} \small The phase diagram corresponding
  to the mean-field Hamiltonian (\ref{HMF}). The shaded region
  schematically denotes the regime where the mapping between liquid
  dynamics and the spin model does not apply. Labels ``Ising magnet''
  and ``Heisenberg magnet'' indicate ordering of the compression and
  shear components in Eq.(\ref{HMF}) in the respective regions of the
  diagram.}
\end{figure}

The Curie temperature itself, according to Eq.(\ref{G}), is determined
by the bigger of the two coupling constants:
\begin{equation} \label{TCurie} k_B T_\sCurie^{\sMF} = \frac{1}{6}
  \mbox{max}(J_\scompr, J_\sshear).
\end{equation}
We remind the reader that the activated regime corresponds to the
ordered state of the magnet.  If $J_\scompr > J_\sshear$, the ordering
is along the compression/dilation component of the local displacement,
which corresponds to the first sum in Eq.(\ref{HMF}), see
Fig.\ref{Ghh}. Note the depths of the two minima are equal, implying
there is no net volume change during the ordering transition.  From
the spin perspective, this state is essentially an ordered Ising
ferromagnet.  Conversely, if $J_\sshear > J_\scompr$, it is the shear
component that becomes ordered; this component corresponds to the
second sum in Eq.(\ref{HMF}).  The frozen-in shear state is an ordered
5-component Heisenberg ferromagnet, from the spin viewpoint.  We
summarize these notions graphically in the phase diagram shown in
Fig.\ref{PhaseDiagram}.

\begin{figure}[t]
\centering 
\includegraphics[width=0.8 \figurewidth]{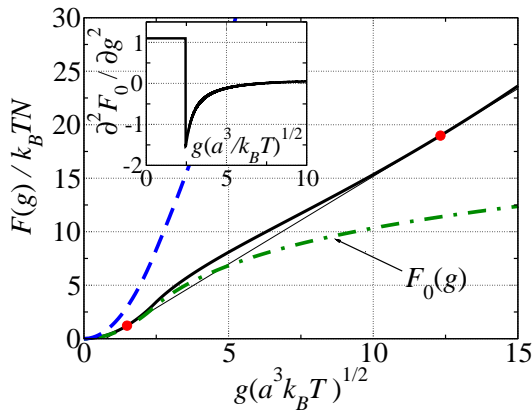}
\caption{\label{FgMF} \small The dashed-dotted line shows the
  mean-field expression for the $F_0(g)$ function. The thick solid
  line shows the full free energy from Eq.(\ref{Ffull}), with a
  specific choice of the function $F_\sex(g)/N = c (g^2 a^3)$, where
  the coefficient $c = 0.05$. The thin solid line is the common
  tangent of the two portions on the $F(g)$ that correspond to the
  mechanically stable reference state and the aperiodic metastable
  state. The dashed line shows the curve with the largest value of
  this coefficient ($c = 5/3$), at which the $F(g)$ curve still
  exhibits a non-concave portion. In the inset, we plot the second
  derivative of the mean-field $F_0(g)$. The discontinuity is at $g^2
  a^3/k_B T = 6$, c.f. Eq.(\ref{gTCurie}).}
\end{figure}

We may further connect the mean-field parameters $J_\sshear$ and
$J_\scompr$ to the material constants by enforcing the aforementioned
notion that the large $g$ asymptotic of the function $F_0(g)$ be
logarithmic in the large $g$ limit, i.e. when the vector length is
large.  According to Eqs.(\ref{Fphig}) and (\ref{G}), the quadratic
term in $g$ from Eqs.(\ref{Fphig}) will cancel out at large $g$, if
\begin{equation} \label{Jg}
\mbox{max}(J_\scompr, J_\sshear) = g^2 a^3. 
\end{equation}
As a result,
\begin{equation} \label{gTCurie} g^2 = \frac{6 k_B
    T^{\sMF}_\sCurie}{a^3}.
\end{equation}
After comparing this equation with Eq.(\ref{gcrit}), we conclude that
the mean-field Curie temperature of our magnet is significantly higher
than the temperature at which the aperiodic stress pattern sets in,
since the actual value of $g$ in the metastable phase is significantly
greater than the critical value from Eq.(\ref{gcrit}), see
Fig.\ref{FgMF}. This notion confirms that, at least in the mean-field
limit, our mapping activated transport onto an ordered spin model is
internally consistent.

We further use the relation in Eq.(\ref{Jg}) to compute the mean-field
expression for $F_0(g)$ (by numerically minimizing the free energy
from Eq.(\ref{G})). The result is shown with the dash-dotted line in
Fig.\ref{FgMF}. To compute the full free energy of the solid $F(g)$,
from Eq.(\ref{Ffull}), we must make a specific assumption on the
penalty $F_\sex(g)$ for dilating the sample. For the sake of argument,
we have used a quadratic function $F_\sex(g)/Na^3 = c g^2$. A specific
value $c = 0.05$ was chosen, so that the resulting barrier between the
two phases is about $k_B T$, since it is known that the one-particle
barrier for surface melting is about $k_B T$.\cite{L_Lindemann}

The mean-field solution supports the asymptotic analysis from Section
\ref{Glass} that the fully mechanically stable solid is indeed the
lowest free energy state; whereas the aperiodic state, though
significantly higher in free energy, is metastable owing to the
surface tension between the two phases.

\section{Summary and Discussion}

We have mapped the liquid activated-transport regime onto the dynamics
of a spin model on a {\em fixed} lattice. The long-lived aperiodic
arrangements, which are characteristic of the activated regime,
correspond to aperiodic spin configurations below the Curie point of
the spin system. The presence of non-zero shear resistance in the
long-lived structures is reflected in the spins having six components,
which correspond to the six independent entries of the local
deformation tensor. Each spin corresponds to a rigid, weakly
interacting molecular units (or ``bead'') of the ROFT theory.  The
size of the bead is set unambiguously as the length scale of the
inhomogeneity of a frozen-in stress distribution in the activated
regime. The length of the vectors is determined self-consistently as
the value of the order parameter for the periodic-to-aperiodic crystal
transition. The mutual angles between spins change little during the
cooperative reconfigurations, implying individual bonds are deformed
nearly harmonically most of the time. The anharmonicity of the
reconfigurations is reflected in a high nucleation barrier in the free
energy profile of the transition, the progress coordinate being the
number of beads that have already switched their positions to their
value in the new arrangement. The present approach exhibits clear
parallels with earlier, self-consistent phonon theories of aperiodic
crystals,\cite{dens_F1, ISI:A1987G269600055} which considered the
uniform liquid state as the reference state and have demonstrated that
the uniform liquid can cross-over to an aperiodic-crystal state,
accompanied by a discontinuous change of an order parameter
corresponding to the bead localization length. In contrast, here we
construct the aperiodic crystal using a fully mechanically-stable
state as the starting point, thus allowing for a straightforward
treatment of local, high-frequency shear resistance. We have also
explicitly demonstrated that aperiodic crystals exhibit multiply
degenerate states whose mutual interconversions can be mapped onto the
dynamics of long-range, 6-spin model with anisotropic interactions.

The present microscopic picture has implications for the temperature
dependence of the activated barriers in supercooled melts that we can
begin to discuss already at the mean-field level. In the allowed part
of the phase diagram in Fig.\ref{PhaseDiagram}, consider a transition
from the $J_\scompr < J_\sshear$ sector to the $J_\scompr > J_\sshear$
sector. As already mentioned, this transition would correspond to
ordering of an Ising-like assembly of spins and and vice versa for the
5-component Heisenberg spins. To avoid confusion, we emphasize that
this Ising-like ordering takes place along one component of randomly
oriented, {\em six}-component vectors. Now, according to accurate
calculations,\cite{PhysRevB.62.9599} the heat capacity jump upon
ordering in the classical Heisenberg model tends to increase with
lowering the dimensionality of the spins and reaches its largest value
in the Ising model. This implies that the r.h.s. part of the phase
diagram should generically exhibit a higher specific heat. In view of
the RFOT-derived connection between the fragility index and the heat
capacity jump, i.e. the mentioned $m \simeq 34.7 \Delta c_p$ relation,
we may expect that substances with a larger $J_\scompr/J_\sshear$
ratio will generically be more fragile, but under several, rather
restrictive circumstances, as we discuss next.

First, we reiterate the present approach is a minimalist way to
explicitly account for the interactions that give rise to (high
frequency) shear resistance in supercooled liquids.  Already in this
minimal model, the entropy and heat capacity per bead depend on the
bead size $a$, which is, formally, the ultraviolet cut-off in the
theory. In addition to the purely volumetric affect - one vector
occupies a volume $a^3$ - the bead size will affect the precise,
self-consistently determined value of the order parameter $g$ in the
glass phase. Further, the isotropic assumption for local elastic
response, from Eq.(\ref{Liso}), is an approximation. Contributions
other than purely elastic terms in Eq.(\ref{fr}) will be generally
present as well.  One source of such contribution is
reconfiguration-induced electric dipole moment. Earlier
estimates\cite{LSWdipole} suggest that even though the local
polarization resulting from bead movements contributes only about a
percent to local elastic constants, the resulting elemental electric
dipoles interact comparably strongly to the elastic dipoles of the
type we have considered. This effect should be especially significant
in ionic glasses.\cite{Lionic} Interestingly, because of the mentioned
antiferromagnetism of the electric dipoles, the electric and elastic
interactions may be mutually frustrating. A careful treatment of
polymeric materials, on the other hand, should include the effects of
chain rigidity and other types of local anisotropy that could not be
accounted for by only two elastic constants and the bead size $a$,
which are the parameters of our ansatz. Novikov and Sokolov have
argued electronic contributions must be considered in metallic
glasses.\cite{NovikovSokolovPRB2006}

According to Fig.\ref{PhaseDiagram}, we should expect that actual
substances will exhibit a continuous range of local stress
distributions and heat capacities. Indeed, to distinct points on the
phase diagram in Fig.\ref{PhaseDiagram}, there correspond very
different values of the heat capacity. Furthermore, the heat capacity
of an anisotropic Heisenberg model on a periodic lattice should
exhibit a non-monotonic behavior as a function of anisotropy, namely a
spike at a transition that would occur at $J_\scompr = J_\sshear$ in
the mean-field limit.  $J_\scompr = J_\sshear$ corresponds to the
following value of the ratio of the longitudinal and transverse
velocities: $c_l/c_t \simeq 1.63$. This is roughly consistent with a
very broad range of fragilities observed for substances from a
relatively narrow range of the $c_l/c_t$ ratio centered at $1.75$ or
so.\cite{Johari_poisson} We point out that the correlation between the
Poisson ratio and the fragility has been a subject of debate for some
time.\cite{PoissonFrag, NovikovSokolovPRB2006, NovikovSokolovKisliuk,
  KNelson_fragility, Johari_poisson, Johari_fragility} Our results
suggest that a correlation between the fragility and the Poisson ratio
might in fact be expected for non-metallic substances that are far
from the transition region $J_\scompr = J_\sshear$. Yet for substances
near the ``critical region'' $J_\scompr = J_\sshear$, little
correlation is expected.  Note that the view of the activated regime
in a {\em fragile} substance as a random Ising model below symmetry
breaking is consistent with recent results of Stevenson at
el.\cite{stevenson:194505} who have mapped the localization transition
in fiducial liquid structures of a Lennard-Jones mixture onto a
replica-symmetry breaking transition in a random Ising model.

Regardless of the detailed value of the heat capacity, our results
indicate that supercooled liquids exhibit local stress distribution
ranging from frozen-in compression to frozen-in shear.  Furthermore,
the present findings that the activated liquid regime arises
self-consistently from a mechanically stable reference state merge
nicely with the self-consistent phonon view of the emergence of the
aperiodic crystal state from the uniform {\em liquid}.\cite{dens_F1,
  ISI:A1987G269600055} This notion suggests that a unified,
quantitative treatment of the liquid, aperiodic-crystal, and
periodic-crystal regimes is in sight. We conclude by reiterating that
despite system-specific variations in direct interactions, accounted
for, semi-empirically, in the variation of the elastic constants in
Fig.\ref{PhaseDiagram}, the underlying mechanism of activated
transport in non-polymeric liquids is system-independent, consistent
with the conclusions of the RFOT theory.

{\bf Acknowledgments}: 
The authors gratefully acknowledge the Arnold and Mabel Beckman
Foundation Beckman Young Investigator Award, the Donors of the
American Chemical Society Petroleum Research Fund, and the Small Grant
and GEAR Programs at the University of Houston, for
partial support of this research. \\


\appendix

\section{Voigt-Mandel notation}
\label{Voigt}

It is sometimes convenient to present the six independent entries of a
symmetric tensor $u_{ij}$ as a six-component vector:
\begin{equation}
  \label{eq:Voigt.vector}
  \{u_{ij}| i,j=1,2,3\} \rightarrow \vec{t} \equiv
  \lp
\begin{array}{c}
  u_{11}\\
u_{22}\\
u_{33}\\
\sqrt{2}u_{23}\\
\sqrt{2}u_{31}\\
 \sqrt{2}u_{12} 
\end{array} 
\rp,  \vspace{3mm}
\end{equation}
so that the convolution $u_{ij} \, \Lambda_{ijkl} \, u_{kl}$ is
expressed as bilinear form $\sum_{\alpha, \beta = 1}^6 t_\alpha
\Lambda_{\alpha \beta} t_\beta$, where $\Lambda_{\alpha \beta}$ is a
square $6\times6$ matrix:
\begin{equation} \label{eq:Voigt.matrix}
\hspace{-1mm}  
{\tiny
    \left(
\begin{array}{cccccc}
  \Lambda_{1111}&\Lambda_{1122}&\Lambda_{1133}&\sqrt{2}\Lambda_{1123}&\sqrt{2}\Lambda_{1131}&\sqrt{2}\Lambda_{1112}\\
  \Lambda_{2211}&\Lambda_{2222}&\Lambda_{2233}&\sqrt{2}\Lambda_{2223}&\sqrt{2}\Lambda_{2231}&\sqrt{2}\Lambda_{2212}\\
  \Lambda_{3333}&\Lambda_{3322}&\Lambda_{3333}&\sqrt{2}\Lambda_{3323}&\sqrt{2}\Lambda_{3331}&\sqrt{2}\Lambda_{3312}\\
  \sqrt{2}\Lambda_{2311}&\sqrt{2}\Lambda_{2322}&\sqrt{2}\Lambda_{2333}&2\Lambda_{2323}&2\Lambda_{2331}&2\Lambda_{2312}\\
  \sqrt{2}\Lambda_{3111}&\sqrt{2}\Lambda_{3122}&\sqrt{2}\Lambda_{3133}&2\Lambda_{3123}&2\Lambda_{3131}&2\Lambda_{3112}\\
  \sqrt{2}\Lambda_{1211}&\sqrt{2}\Lambda_{1222}&\sqrt{2}\Lambda_{1233}&2\Lambda_{1223}&2\Lambda_{1231}&2\Lambda_{1212}
\end{array}
\right) \!\!.} \! \vspace{3mm}
\end{equation}

In the isotropic case, see Eq.(\ref{Liso}), this matrix is rather
sparse and easily shown to have one eigenvalue $3K$ and a five-fold
degenerate eigenvalue $2\mu$.  Importantly, the corresponding
orthogonal transformation matrix, i.e the matrix consisting of the
eigenvectors of the matrix from Eq.(\ref{eq:Voigt.matrix}) for an
isotropic $\Lambda$, {\em does not depend on the $K/\mu$ ratio}. (The
transformation matrix is easy to compute, but is too bulky to give
here.) This means, among other things, that in the Voigt basis, {\em
  all} isotropic interactions are diagonal.

One may further rescale the components of the 6-vectors, for a
specific combination of $K$ and $\mu$:
\begin{equation} \label{eq:ttod.explicit} \vec{g}=
  \left( \begin{array}{c}
      \sqrt {K}\left(d_{11}+d_{22}+ d_{33} \right) \\
      \noalign{\medskip}\sqrt{\mu} \left(d_{11}- d_{22}\right) \\
      \noalign{\medskip}\sqrt{\frac{\mu}{3}}\left(d_{11}+
        d_{22}-2d_{33}
      \right) \\
      \noalign{\medskip}2\sqrt{\mu}d_{23} \\
      \noalign{\medskip}2\sqrt{\mu}d_{31} \\
      \noalign{\medskip}2\sqrt{\mu}d_{12}
    \end{array} \right) 
\end{equation}
so as to make the matrix from Eq.(\ref{eq:Voigt.matrix}) not only
diagonal, but also unit: $d_{ij} \Lambda_{ijkl} d_{kl} = \vec{g} \, ^2
\equiv \sum_{\alpha=1}^6 g_\alpha^2 \equiv g^2$ (and similarly for the
elastic components), see Eqs.(\ref{phidef}-\ref{gdef}).

\section{Spin-spin interaction}
\label{K}

To evaluate the coupling between the non-linear oscillators, we
integrate out only the phonon degrees of freedom $\bu$ in
Eq.(\ref{Fug}), i.e.  we need to perform the following functional
integral:
\begin{equation} \label{fr_short} \int [\Pi_{\br} d^3 \bu(\br)] e^{
    -\beta [\int d^3\br ( u_{ij} \Lambda_{ijkl} u_{kl}/2) + \sum_m a^3
    d_{ij}(\br_m) \Lambda_{ijkl} u_{kl}(\br_m)] },
\end{equation}
where the $m$-summation is over the locations $\br_m$ of the harmonic
oscillators. (Note the $d \Lambda u$ cross-term is symmetrized because
$\Lambda_{ijkl} = \Lambda_{klij}$.) In terms of the Fourier components
of the local displacements $\bu$: $ \btu(\bk) = \int d^3 \br \:
\bu(\br) e^{i \bk \br} $, this integral becomes:
\begin{eqnarray}
  \label{fr_shortK} \int & \hspace{-4mm} [\Pi_{\bk} d^3 \btu(\bk)] \exp\left\{
    -\beta \int \frac{d^3\bk}{(2 \pi)^3} [ \Lambda_{njlm} \tu_n(\bk)
    \tu_l(-\bk) k_j k_m/2 \right. \nonumber \\ 
  &+ \left. \sum_m a^3 d_{nj}(\br_m) \, \Lambda_{njlp}
    (-ik_l) \tu_{p} e^{-i \bk \br_m}] \right\},
\end{eqnarray}
where an appropriate ultraviolet cut-off at $k_\smicro$, as in
Eq.(\ref{Kijkl}), is understood. (Note $\Lambda_{ijkl} =
\Lambda_{jikl}$, $\Lambda_{ijkl} = \Lambda_{ijlk}$.) For the specific,
simple form of the elastic tensor $\Lambda$ from Eq.(\ref{Liso}), it
is straightforward to compute the Gaussian integral above explicitly,
using the obvious property of the Fourier component of the real-valued
displacements $\bu$: $\btu(-\bk) = \btu^*(\bk)$, and
Hubbard-Stratonovich formulas $\int_{-\infty}^{\infty} \frac{dx
  dy}{\pi} \exp[-(x^2 + y^2) + a (x-iy) + b^*(x+iy)] = e^{a b^*}$ and
$\int_{-\infty}^\infty (\Pi_m dx_m/\pi^{1/2}) e^{-x_k A_{kl} x_l + b_k
  x_k} = (\mbox{Det} A)^{-1/2}e^{b_k A^{-1}_{kl} b_l/4}$, where
$\{A_{kl}\}$ is a positively defined symmetric matrix.  The inverse of
the matrix $A_{il} \equiv \Lambda_{ijlm} k_j k_m $ is easy to find,
using the special property of the matrix $P_{jm} \equiv k_j k_m/k^2
\equiv \hk_j \hk_m$ that $P^2 = P$. Using this notion, one may show
that
\begin{equation} \label{Linverse} (\Lambda_{ijlm} \hk_j \hk_m)
  \left(\delta_{ln} - \frac{\lambda +\mu}{\lambda+2\mu} \hk_l \hk_n
  \right) = \delta_{in}.
\end{equation}
where $\Lambda_{ijkl}$ is from Eq.(\ref{Liso}). Finally note that the
$k$-integration in the exponent in Eq.(\ref{fr_shortK}) counts every
phononic mode twice, because of the mentioned property $\btu(-\bk) =
\btu^*(\bk)$. ($d^3 \btu(\bk) \equiv \{ d^3 \mbox{Re}[\btu(\bk)] d^3
\mbox{Im}[\btu(\bk)] \}^{1/2}$.)  We use the listed notions to perform
the functional integration in Eq.(\ref{fr_shortK}) and thus arrive at
Eq.(\ref{H}) of the main text.

It may be instructive to compare the straightforward, if not somewhat
tedious calculation above with a simpler case of interaction between
3-component vectors mediated by a single-polarization elastic
interaction:
\begin{equation} E = \int dV [K({\bm \nabla} \psi)^2/2 + \sum_m
  (\bg^{(m)} {\bm \nabla} \psi) \delta^3(\br-\br_m)],
\end{equation}
where $\psi \equiv \psi(\br)$ is a coordinate dependent, scalar field.
In this case, the phonon-mediated coupling between the spins becomes:
\begin{equation} {\cal H} = - \sum_{m<n} a^3 J_{ij}(\br_{mn})
  g_i^{(m)} g_j^{(n)},
\end{equation}
This coupling looks particularly simple in the Fourier domain:
$J_{ij}(\br_{mn}) = \int \frac{d^3(k a)}{(2\pi)^3} \cos(\bk \br) \hk_i
\hk_j$; at distances well-exceeding the inverse ultraviolet cut-off,
it is identical to the usual electric dipole-dipole interaction,
albeit with the opposite sign: $J_{ij}(\br_{mn}) \equiv (\delta_{ij} -
3 n_i n_j)/r^3_{mn}$, where $\br_{mn} \equiv (\br_m - \br_n)$ and $\bn
\equiv \br_{mn}/r_{mn}$. It is easy to show, by a direct calculation,
that the eigenvalues of the matrix $\hk_i \hk_j$ are 1, 0, 0.

To perform the angular averaging of the tensor $\tilde{K}$ in
Eq.(\ref{Kijkl}), one needs to compute $\langle \hk_i \hk_j \rangle$
and $\langle \hk_i \hk_j \hk_l \hk_m \rangle$, where the angular
brackets $\langle \ldots \rangle$ denote angular averaging. To compute
these averages, we note that by symmetry, $\langle \hk_i \hk_j \rangle
\propto \delta_{ij}$ and that $\hk_i^2 =1$, implying $\langle \hk_i
\hk_j \rangle = (1/3) \delta_{ij}$. By the same token, since $\langle
\hk_i \hk_j \hk_l \hk_m \rangle \propto (\delta_{ij} \delta_{lm} +
\delta_{il} \delta_{jm} + \delta_{im} \delta_{jl})$ and $\hk_i^2
\hk_j^2 =1$, we obtain $\langle \hk_i \hk_j \hk_l \hk_m \rangle =
(1/15) (\delta_{ij} \delta_{lm} + \delta_{il} \delta_{jm} +
\delta_{im} \delta_{jl})$. Using these expressions and
Eq.(\ref{Liso}), we obtain $\langle \tilde{K}_{ijkl} \rangle =
\lambda' \delta_{ij} \delta_{kl} + \mu' (\delta_{ik} \delta_{jl} +
\delta_{il} \delta_{jk})$, where $\lambda' = (\lambda^2 + 16\lambda
\mu/15 - 4\mu^2/15)/(\lambda + 2\mu)$ and $\mu' = 2 \mu (3\lambda + 8
\mu)/15(\lambda + 2\mu)$. Therefore the $\langle \tilde{K}_{ijkl}
\rangle$ tensor has the same structure as the $\Lambda$ tensor itself
and will be diagonal in the Voigt representation, after a basis
change as in Eq.(\ref{eq:ttod.explicit}). Of the corresponding
diagonal entries, one is $K/(\lambda + 2 \mu)$ and the rest five are
$(2/5)(K+2\mu)/(\lambda+2\mu)$, as mentioned in the main text. We note
that the sum of the diagonal entries is equal to 3, as expected from
the discussion in the beginning of Section \ref{MF}.
\bibliographystyle{jpc}
\bibliography{lowT}
\end{document}